\documentclass[useAMS,usenatbib]{mn2e}
\input psfig.sty
\usepackage{flafter}

\def \gas {\textsc{gasoline}}
\def \gastwo {\textsc{ESF-gasoline2}}

\def \apj {ApJ}
\def \apjl {ApJL}
\def \mnras {MNRAS}
\def \apjs {ApJS}
\def \aap {A\&A}

\def \pasp {PASP}
\def \na {NewA}

\def \etal {et~al.~}

\def \spose#1{\hbox  to 0pt{#1\hss}}  
\def \lta{\mathrel{\spose{\lower 3pt\hbox{$\sim$}}\raise  2.0pt\hbox{$<$}}}
\def \gta{\mathrel{\spose{\lower  3pt\hbox{$\sim$}}\raise 2.0pt\hbox{$>$}}}

\def \kms {\ifmmode  \,\rm km\,s^{-1} \else $\,\rm km\,s^{-1}  $ \fi }
\def \kpc {\ifmmode  {\rm kpc}  \else ${\rm  kpc}$ \fi  }  
\def \hkpc {\ifmmode  {h^{-1}\rm kpc}  \else ${h^{-1}\rm kpc}$ \fi  }  
\def \hMpc {\ifmmode  {h^{-1}\rm Mpc}  \else ${h^{-1}\rm Mpc}$ \fi  }  
\def \Msun {\ifmmode \rm M_{\odot} \else $\rm M_{\odot}$ \fi}
\def \hMsun {\ifmmode h^{-1}\,\rm M_{\odot} \else $h^{-1}\,\rm M_{\odot}$ \fi}
\def \hhMsun {\ifmmode h^{-2}\,\rm M_{\odot}\else $h^{-2}\,\rm M_{\odot}$ \fi}
\def \Lsun {\ifmmode L_{\odot} \else $L_{\odot}$ \fi} 
\def \hhLsun {\ifmmode h^{-2}\,\rm L_{\odot} \else $h^{-2}\,\rm L_{\odot}$ \fi}

\def\LCDM{$\Lambda$CDM }
\def \LCDM {\ifmmode \Lambda{\rm CDM} \else $\Lambda{\rm CDM}$ \fi}
\def \sig8 {\ifmmode \sigma_8 \else $\sigma_8$ \fi} 
\def \Omegam {\ifmmode \Omega_{\rm m} \else $\Omega_{\rm m}$ \fi} 
\def \Omegab {\ifmmode \Omega_{\rm b} \else $\Omega_{\rm b}$ \fi} 
\def \Omegar {\ifmmode \Omega_{\rm r} \else $\Omega_{\rm r}$ \fi} 
\def \fbar {\ifmmode f_{\rm bar} \else $f_{\rm bar}$ \fi} 
\def \OmegaL {\ifmmode \Omega_{\rm \Lambda} \else $\Omega_{\rm \Lambda}$\fi} 
\def \Deltavir {\ifmmode \Delta_{\rm vir} \else $\Delta_{\rm vir}$ \fi}
\def \rhocrit {\ifmmode \rho_{\rm crit} \else $\rho_{\rm crit}$ \fi}

\def \rs {\ifmmode r_{\rm s} \else $r_{\rm s}$ \fi} 
\def \rh {\ifmmode r_{\rm h} \else $r_{\rm h}$ \fi} 
\def \Rvir {\ifmmode R_{\rm vir} \else $R_{\rm vir}$ \fi}
\def \Vvir {\ifmmode V_{\rm  vir} \else  $V_{\rm vir}$  \fi} 
\def \Vmax {\ifmmode V_{\rm  max} \else  $V_{\rm max}$  \fi} 
\def \Mvir {\ifmmode M_{\rm  vir} \else $M_{\rm  vir}$ \fi}  
\def \Mhalo {\ifmmode M_{\rm halo} \else $M_{\rm  halo}$ \fi}  
\def \Nvir {\ifmmode N_{\rm  vir} \else $N_{\rm  vir}$ \fi}  
\def \Jvir {\ifmmode J_{\rm vir} \else $J_{\rm vir}$ \fi} 
\def \Evir {\ifmmode E_{\rm vir} \else $E_{\rm vir}$ \fi} 
\def \lam {\ifmmode \lambda  \else $\lambda$ \fi} 
\def \lamp {\ifmmode \lambda^{\prime} \else $\lambda^{\prime}$  \fi} 
\def \lampc {\ifmmode \lambda^{\prime}_{\rm c} \else
  $\lambda^{\prime}_{\rm c}$  \fi} 

\def \xoff {\ifmmode x_{\rm off} \else $x_{\rm off}$ \fi}
\def \rhorms {\ifmmode \rho_{\rm rms} \else $\rho_{\rm rms}$ \fi}
\def \qbar {\ifmmode \bar{q} \else $\bar{q}$ \fi}

\def \Mb {\ifmmode M_{\rm b} \else $M_{\rm b}$ \fi} 
\def \eSF {\ifmmode \epsilon_{\rm SF} \else $\epsilon_{\rm SF}$ \fi} 
\def \Md {\ifmmode M_{\rm d} \else $M_{\rm d}$ \fi} 
\def \Mg {\ifmmode M_{\rm g} \else $M_{\rm g}$ \fi} 
\def \Rb {\ifmmode R_{\rm b} \else $R_{\rm b}$ \fi} 
\def \Rd {\ifmmode R_{\rm d} \else $R_{\rm d}$ \fi} 
\def \Rg {\ifmmode R_{\rm g} \else $R_{\rm g}$ \fi} 
\def \mgal {\ifmmode m_{\rm gal} \else $m_{\rm gal}$ \fi} 
\def \rj {\ifmmode {\cal R}_j \else ${\cal R}_j$ \fi} 
\def \lamgal {\ifmmode \lambda_{\rm gal} \else $\lambda_{\rm gal}$ \fi} 
\def \Vcirc {\ifmmode V_{\rm circ} \else $V_{\rm circ}$ \fi} 
\def \Vrot {\ifmmode V_{\rm rot} \else $V_{\rm rot}$ \fi} 
\def \Vflat {\ifmmode V_{\rm flat} \else $V_{\rm flat}$ \fi} 
\def \Mstar {\ifmmode M_{\rm star} \else $M_{\rm star}$ \fi} 
\def \Mgas {\ifmmode M_{\rm gas} \else $M_{\rm gas}$ \fi} 
\def \Mbar {\ifmmode M_{\rm bar} \else $M_{\rm bar}$ \fi} 

\def \DeltaIMF {\ifmmode \Delta_{\rm IMF} \else $\Delta_{\rm IMF}$ \fi}

\def \VV {\ifmmode V_{\rm 2.2}/V_{200} \else $V_{2.2}/V_{200}$ \fi} 
\def \dvr {\ifmmode \partial_{\rm VR} \else $\partial_{\rm VR}$ \fi} 

\title[Reproducing inefficient galaxy formation] {NIHAO project I:
  Reproducing the inefficiency of galaxy formation across cosmic
  time with a large sample of cosmological hydrodynamical simulations.}

\author[Wang et al.]{Liang Wang$^{1,2,3}$\thanks{lwang@mpia.de}, Aaron A. Dutton$^2$,
  Gregory S. Stinson$^2$, Andrea V. Macci\`o$^2$, 
\newauthor{Camilla Penzo$^1$, Xi Kang,$^1$ Ben W. Keller$^4$,
  James Wadsley$^4$}\\
$^1$Purple Mountain Observatory, the Partner Group of MPI f\"ur Astronomie, 2 West Beijing Road, Nanjing 210008, China\\
$^2$Max-Planck-Institut f\"ur Astronomie, K\"onigstuhl 17, 69117 Heidelberg, Germany\\
$^3$University of Chinese Academy of Science, 19A Yuquan Road, 100049 Beijing, China\\
$^4$Department of Physics and Astronomy, McMaster University, Hamilton, Ontario L8S 4M1, Canada}
\begin{document}

\date{submitted to MNRAS}
             
\pagerange{\pageref{firstpage}--\pageref{lastpage}}\pubyear{2015}

\maketitle           

\label{firstpage}
             

\begin{abstract}
  We introduce project {\sc NIHAO} (Numerical Investigation of a
  Hundred Astrophysical Objects), a set of 100 cosmological zoom-in
  hydrodynamical simulations performed using the {\sc gasoline} code,
  with an improved  implementation of the SPH algorithm.  The haloes
  in our study range from dwarf ($M_{200} \sim 5\times10^{9}\Msun$) to
  Milky Way ($M_{200}\sim 2\times 10^{12}\Msun$) masses, and represent an
  unbiased sampling of merger histories, concentrations and spin
  parameters. The particle masses and force softenings are chosen to
  resolve the mass profile to below 1\% of the virial radius at all
  masses, ensuring that galaxy half-light radii are well
  resolved. Using the same treatment of star formation and stellar
  feedback for every object,  the simulated galaxies reproduce the
  observed inefficiency of galaxy formation across cosmic time as
  expressed through the stellar mass vs halo mass relation, and the
  star formation rate vs stellar mass relation. We thus conclude that
  stellar feedback is the chief piece of physics required to limit the
  efficiency of star formation in galaxies less massive than the Milky
  Way.  
\end{abstract}


\setcounter{footnote}{1}


\section{Introduction}
\label{sec:intro}

Understanding the formation of spiral galaxies (including our own
Galaxy -- the Milky Way) has been at the forefront of theoretical
astrophysics for decades. While there is a broadly-sketched paradigm
for the formation of disk galaxies (White \& Rees 1978; Fall \&
Efstathiou 1980; Mo, Mao, \& White 1998; Dutton \etal 2007), the
details have yet to be understood.  The simulation of realistic spiral
galaxies from cosmological initial conditions has been a formidable
challenge. Early (low-resolution) simulations were plagued by
over-cooling, and angular momentum losses resulting in  compact bulge
dominated galaxies (e.g., Navarro \& Steinmetz 2000). Recently, it has
been shown that  higher spatial resolution coupled to more realistic
models for star formation and stellar feedback can solve these
problems, resulting in spiral galaxies with realistic sizes and bulge
fractions (Guedes \etal 2011; Brook \etal 2012; Aumer \etal 2013;
Marinacci \etal 2014; Crain \etal 2015). 

One of the most fundamental constraints for galaxy formation models is
the relation between the stellar mass of a galaxy, $\Mstar$, and the
mass of its host dark matter halo, $\Mhalo$. This relation encodes the
global efficiency at which galaxies turn gas into stars. In the
context of $\LCDM$, a model that matches the $\Mstar - \Mhalo$ relation
will also reproduce the observed galaxy stellar mass 
function, which has been a benchmark for galaxy formation models for
over a decade (e.g., White \& Frenk 1991, Benson \etal 2003; Crain
\etal 2009; Oppenheimer \etal 2010). An advantage of the \Mstar vs
\Mhalo relation is that single galaxies can be compared, whereas
comparing with the stellar mass function requires a cosmologically
representative volume.

\begin{figure*}
\centerline{
  \psfig{figure=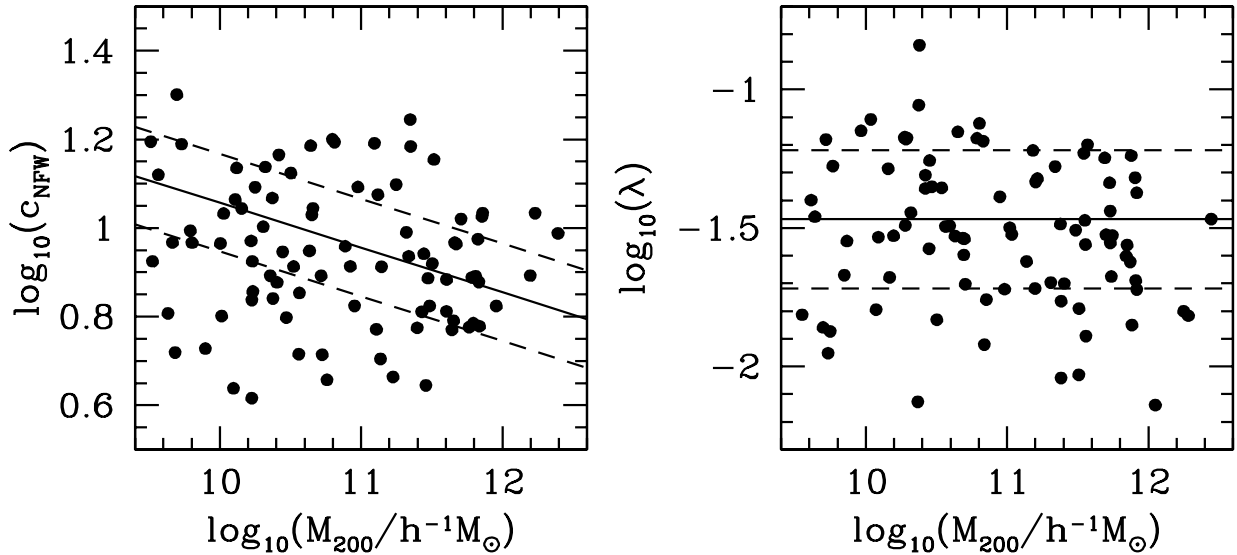,width=0.95\textwidth}
}
\caption{Concentration vs halo mass (left) and spin parameter vs halo
  mass (right) for the sample of haloes presented in this
  paper. Concentrations are from NFW fits. Solid and dashed lines show
  the mean and 1$\sigma$ scatter from the low-resolution parent
  simulations in the Planck cosmology from Dutton \& Macci\`o (2014).}
\label{fig:cm_spin}
\end{figure*}

In recent years the \Mstar vs \Mhalo relation has been studied
extensively using observations of weak gravitational lensing,
satellite kinematics, and halo abundance matching (e.g., Yang \etal
2003, 2012; Mandelbaum \etal 2006;  Conroy \& Wechsler 2009; Moster
\etal 2010;  More \etal 2011; Leauthaud \etal 2012; Behroozi \etal
2013;  Garrison-Kimmel \etal 2014; Kravtsov \etal 2014; Hudson \etal
2015).  One of the key results of these studies is that star formation
is inefficient -- the maximum efficiency is $\sim 25\%$, and
independent of redshift, occurring at a halo mass similar to that of
the Milky Way (i.e., $\sim 10^{12}\Msun$). At both higher and lower
halo masses the efficiency drops even further.

Reproducing this relation both at redshift $z=0$ and earlier times, in
a fully cosmological context has been a challenge, with many
simulations and semi-analytic models predicting either too many stars
at redshift $z=0$, or earlier times (Weinmann \etal 2012).  Even some
of the largest zoom-in and cosmological volume simulations do not
reproduce the observed $z=0$ relation between stellar and halo mass
(e.g., Guedes \etal 2011; Marinnaci \etal 2014; Vogelsberger \etal
2014).  Only a handful of cosmological simulations have been able to
 match the stellar vs halo mass relation over a wide range of halo
masses from dwarfs to the Milky Way (Di Cintio \etal 2014; Hopkins
\etal 2014; Schaye \etal 2015).  On the other hand these simulations
were subject to either limited  resolution on disc galaxy scales
(Schaye \etal 2015) or based on a handful of objects with unclear
selection function (DiCintio \etal 2014; Hopkins \etal 2014) and thus
not able to recover the variety of formation histories of real
galaxies.

The next step is then to create a large sample of galaxies with very
high resolution over a range of halo masses with an unbiased sampling
of the mass accretion histories.  The NIHAO\footnote{Nihao is the
  Chinese word for ``hello''.}  (Numerical Investigation of a Hundred
Astrophysical Objects) project aims to achieve these goals with $\sim
100$ hydrodynamical cosmological zoom-in simulations with halo masses
ranging from $\Mhalo \sim 5\times 10^{9}$ to $\sim 2\times 10^{12} \Msun$.  

In this paper we describe the numerical methods and sample selection
and present results on the growth of stellar mass across cosmic time
from the first 88 NIHAO simulations.  Subsequent papers will
discuss: dark halo shapes (Butsky \etal  2015); central dark matter
density slopes (Tollet \etal 2015); Tully-Fisher relations (Dutton
\etal in prep.); and gas content (Stinson \etal 2015). This paper is
organized as follows: The simulations including sample selection,
hydrodynamics, star formation and feedback are described in
\S\ref{sec:sims}, results including the stellar mass vs halo mass and
star formation rate vs halo mass relation are presented in
\S\ref{sec:results} and a summary in \S\ref{sec:sum}.


\section{Simulations} \label{sec:sims}

The simulations presented here are a series of fully cosmological
``zoom-in'' simulations of galaxy formation run in a flat $\Lambda$CDM
cosmology with parameters from the Planck Collaboration \etal (2014):
Hubble parameter $H_0$= 67.1 \kms Mpc$^{-1}$, matter density
$\Omegam=0.3175$, dark energy density $\Omega_{\Lambda}=1-\Omegam
-\Omegar=0.6824$, radiation density $\Omegar=0.00008$,  baryon density
$\Omegab=0.0490$, power spectrum normalization $\sigma_8 = 0.8344$,
power spectrum slope $n=0.9624$.

\begin{table*}
\begin{center}
  \caption{Dark and gas particle masses, $m$, and force softenings,
    $\epsilon$, for our zoom-in simulations.}
\begin{tabular}{llllll}
\hline
box.level& halo mass range  &$m_{\rm gas}$ & $\epsilon_{\rm gas}$ & $m_{\rm DM}$ & $\epsilon_{\rm DM}$ \\
         & (M$_\odot$) &(M$_\odot$)  & (pc)   &   (M$_\odot$)& (pc) \\
\hline 
  60.1 & $6.96\times 10^{11} - 2.79\times 10^{12}$ & 3.166$\times$10$^5$  & 397.9 & 1.735$\times$10$^6$  & 931.4 \\
  60.2 & $8.89\times 10^{10} - 5.55\times 10^{11}$ & 3.958$\times$10$^4$  & 199.0 & 2.169$\times$10$^5$  & 465.7 \\
  60.3 & $2.63\times 10^{10} - 7.12\times 10^{10}$ & 1.173$\times$10$^4$  & 132.6 & 6.426$\times$10$^4$  & 310.5 \\
  20.1 & $4.99\times 10^{9} - 2.94\times 10^{10}$ & 3.475$\times$10$^3$  & $\phantom{1}$88.4 & 1.904$\times$10$^4$  & 207.0 \\
  15.1 & $5.41\times 10^{9} - 9.26\times 10^{9}$ & 2.087$\times$10$^3$  & $\phantom{1}$74.6 & 1.144$\times$10$^4$  & 174.6 \\
  20.2 & $4.36\times 10^{9}$ & 1.466$\times$10$^3$  & $\phantom{1}$66.4 & 8.032$\times$10$^3$  & 155.2 \\
  15.2 & $3.54\times 10^{9}$ & 6.184$\times$10$^2$  & $\phantom{1}$49.7 & 3.389$\times$10$^3$  & 116.4 \\  
 \hline
\end{tabular}
\label{tab:parameters}
\end{center}
\end{table*}

\subsection{Sample selection and Initial Conditions}

The simulated galaxies were selected from two cosmological boxes of
sides 60 and 20 ${\rm Mpc}/h$ from Dutton \& Macci\`o (2014), as well
as a new simulation of size 15 ${\rm Mpc}/h$ with $400^3$
particles. We select halo masses from $9.5 \lta
\log_{10}(M_{200}/\Msun) \lta 12.3$. Haloes are selected without
considering the halo structure or merger history. For technical
reasons with the zoom-in technique we require the haloes to be
``isolated''. Formally, we only consider haloes that have no other
haloes with mass greater than one-fifth of the virial mass within 3
virial radii. 95\% of all parent haloes satisfy this constraint, so it
does not bias our sample in any significant way. The concentration vs
mass and spin vs mass relations from dark matter only simulations are
shown in Fig.~\ref{fig:cm_spin}. The solid and dashed lines show the
mean and scatter from the parent simulations from Dutton \& Macci\`o
(2014) showing that our selected haloes span a wide range in
concentrations and spins.

Initial conditions for zoom-in simulations were created using a
modified version of the {\sc grafic2} package (Bertschinger 2001) as
described in Penzo \etal (2014).  We chose the refinement level in
order to maintain a roughly constant relative resolution (i.e.,
$\epsilon_{\rm DM}/\Rvir \sim 0.003$), and  $\sim 10^6$ dark matter
particles per halo.  This allows us to reliably probe the mass profile
down to 1\% of the virial radius or better across the full range of
halo masses we simulate. The motivation for achieving a constant
relative resolution, rather than a fixed physical resolution is that
we wish to resolve the dark matter mass profile down to the galaxy
half-light radius, which is typically 1-2\% of the virial radius
(Kravtsov 2013).

\begin{figure}
\centerline{
\psfig{figure=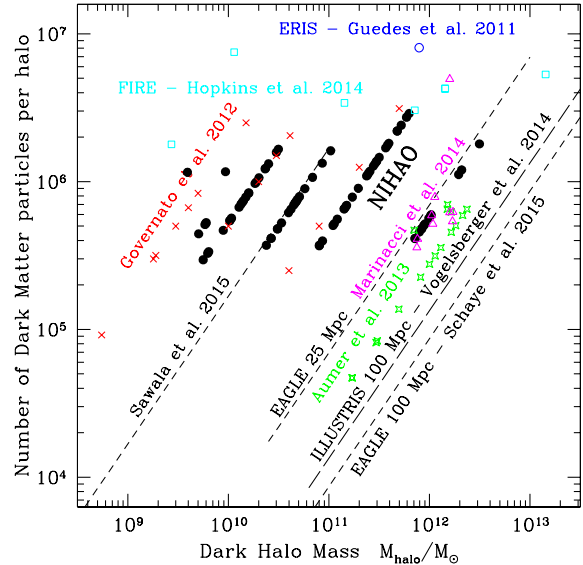,width=0.45\textwidth}
}
\caption{Number of dark matter particles per halo vs halo mass for
  NIHAO simulations (black circles) together with state-of-the-art
  cosmological simulations in the literature.}
\label{fig:massres}
\end{figure}

Fig.~\ref{fig:massres} shows the mass resolution of our simulations
compared to a number of state-of-the-art zoom-in (ERIS - Guedes \etal
2011; Governato \etal 2012; Aumer \etal 2013; Marinacci \etal 2014;
FIRE - Hopkins \etal 2014; Sawala \etal 2015) and large volume (ILLUSTRIS - Vogelsberger
\etal 2014; EAGLE - Schaye \etal 2015) simulations. NIHAO is by far
the largest set of zoom-in simulations resolved with $\sim 10^6$
particles per halo.  The large volume simulations (EAGLE, ILLUSTRIS)
contain many thousands of galaxies, but the fixed force and mass
resolution means that only the highest mass haloes are well resolved.

The force softenings and particle masses for the highest refinement
level for each simulation are given in Table~\ref{tab:parameters}.
Note that the ratio between dark and gas particle masses is initially
the same as the cosmological dark/baryon mass ratio $\Omega_{\rm
  DM}/\Omega_{\rm b}\approx 5.48$, and the gas (and star) particle
force softenings are set to be $\sqrt{(\Omega_{\rm DM}/\Omega_{\rm
    b})}\approx 2.34$ times smaller than the dark matter particle
softenings.  Each hydrodynamic simulation has a corresponding dark
matter-only simulation run at the same mass and force resolution to
enable a study of the effects of galaxy formation on dark halo
structure (Dutton \etal in prep).

\begin{figure*}
\centerline{
\psfig{figure=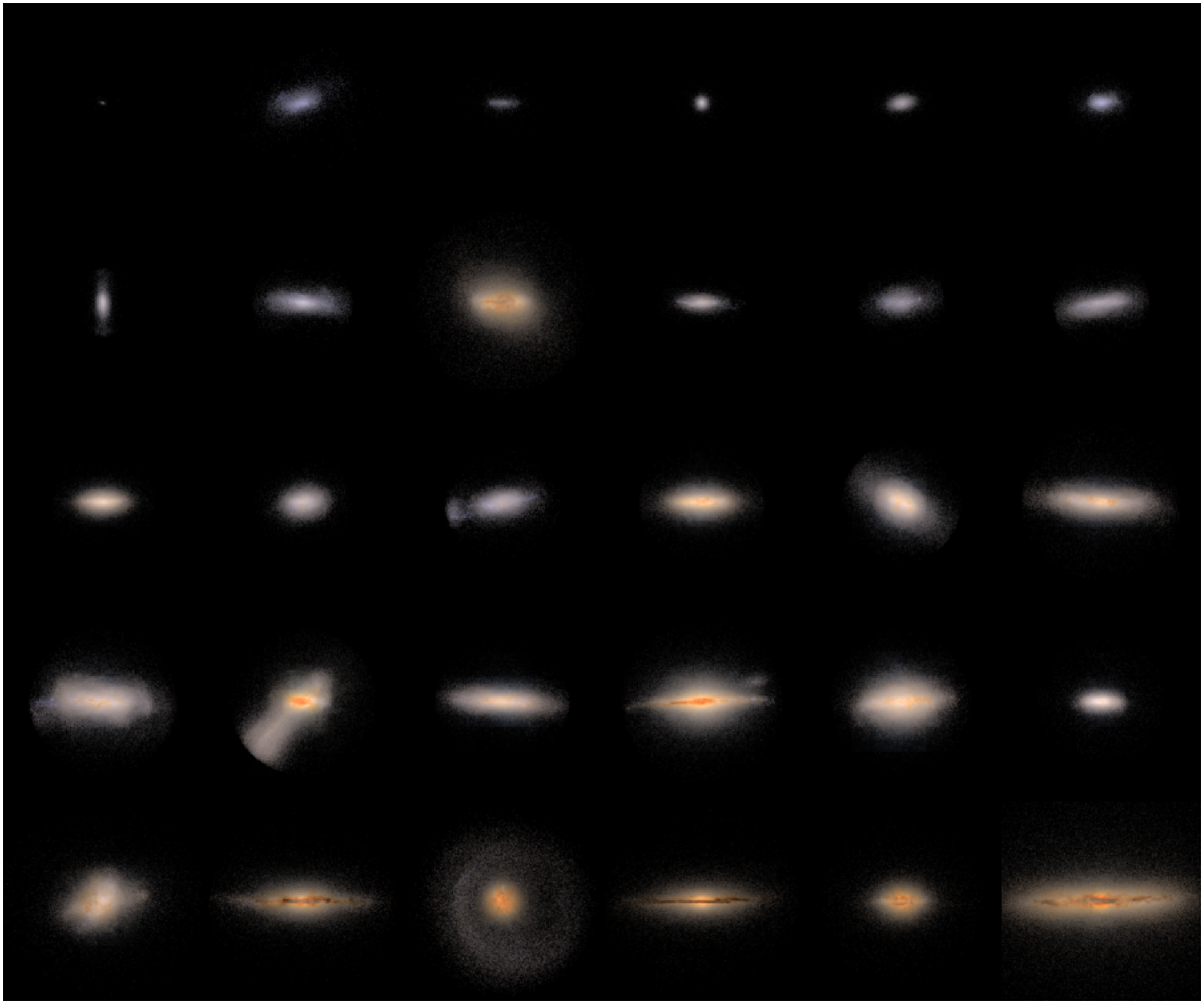,width=0.99\textwidth}
}
\caption{Edge-on views of a subset of NIHAO galaxies after processing
  through the Monte Carlo radiative transfer code
  {\sc{sunrise}}. Images are 50 kpc on a side.}
\label{fig:edgeon}
\end{figure*}

\begin{figure*}
\centerline{
\psfig{figure=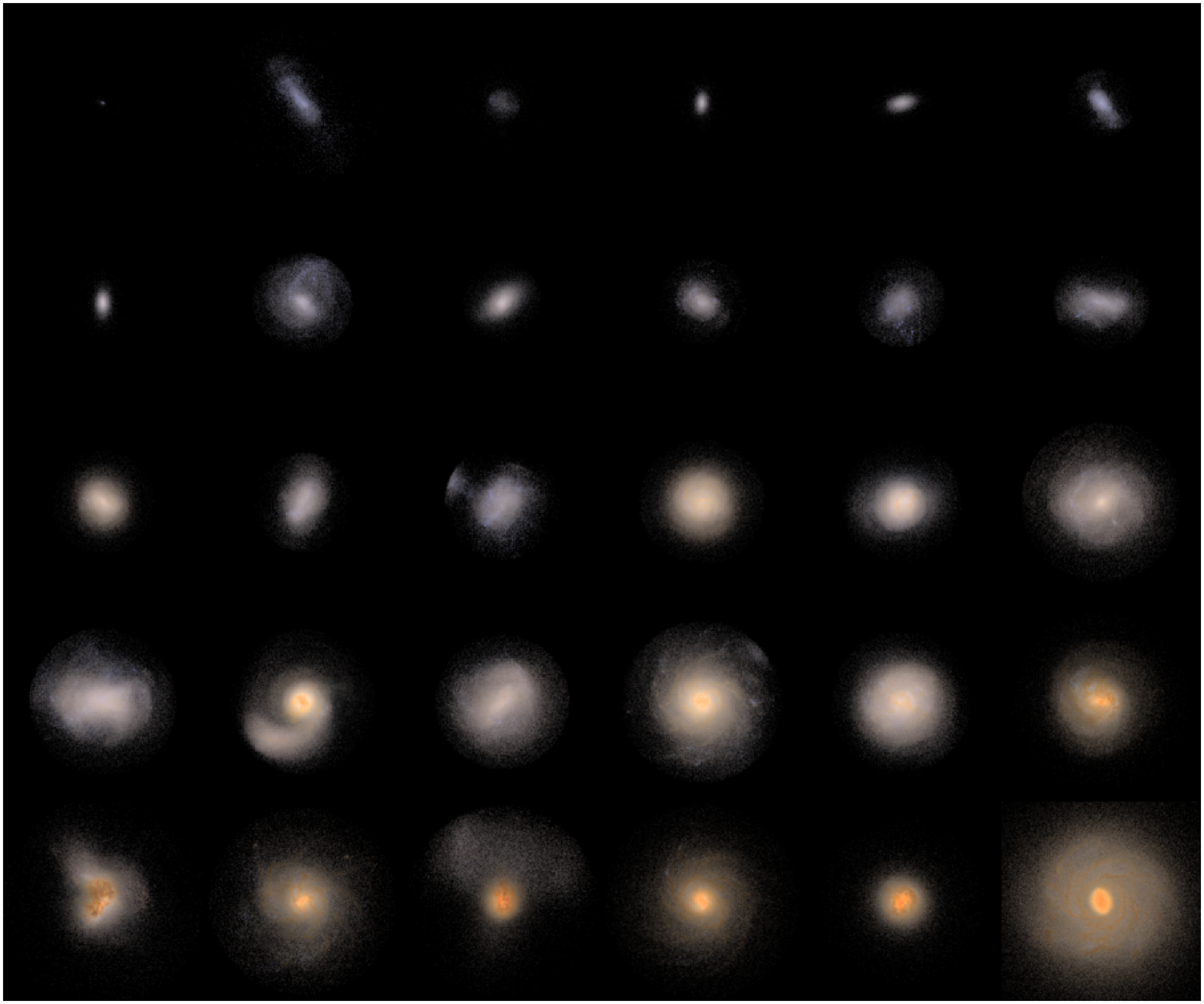,width=0.99\textwidth}
}
\caption{Face-on views of a subset of NIHAO galaxies after processing
  through the Monte Carlo radiative transfer code
  {\sc{sunrise}}. Images are 50 kpc on a side.}
\label{fig:faceon}
\end{figure*}

\subsection{Hydrodynamics}

We use a new version of the N-body SPH solver \textsc{gasoline}
(Wadsley \etal 2004). A complete description of the new gasoline code
including tests is given in Wadsley \etal (in prep.).  In their paper
describing superbubble feedback, Keller \etal (2014) described the
updates they made to {\sc{gasoline}}. While we do not use their
superbubble feedback, we employ their modified version of
hydrodynamics that reduces the formation of blobs and improves mixing.
We thus refer to our version of \textsc{gasoline} as \gastwo.  The
biggest differences to the hydrodynamics in \gastwo\, come from the
small change Ritchie \& Thomas (2001) proposed for calculating
$P/\rho^2$.  Ritchie \& Thomas (2001) also proposed modifying the
density calculation to use equal pressures, but we do \emph{not} use
those densities in the simulations described here.

Diffusion of quantities like metals and thermal energy between
particles has been implemented as described in Wadsley \etal (2008).
Metal diffusion is used, but thermal diffusion is not used because it
is incompatible with the blastwave feedback that delays cooling.
\textsc{Gasoline2} includes several other changes to the hydrodynamic
calculation.  The Saitoh \& Makino (2009) timestep limiter was
implemented so that cool particles behave correctly when a hot
blastwave hits them.  To avoid pair instabilities, \gastwo\, uses the
Wendland $C2$ function for its smoothing kernel (Dehnen \& Aly 2012).
The treatment of artificial viscosity has been modified to use
the signal velocity as described in Price (2008). We also increase the 
number of neighbor particles used in the calculation of the smoothed 
hydrodynamic properties from 32 to 50.

Cooling via hydrogen, helium, and various metal-lines in a uniform
ultraviolet ionizing background is included as described in Shen \etal
(2010) and was calculated using \textsc{cloudy} (version 07.02;
Ferland \etal 1998) These calculations include photo ionization and
heating from the Haardt \& Madau (private communication)  UV
background and Compton cooling and range from 10 to $10^9$ K.
In the dense interstellar medium gas the extragalactic UV field is a
  reasonable approximation, and thus we do not impose any shielding.

\subsection{Star Formation and Feedback}
\label{sec:feedback}
Within the hydrodynamic simulations, gas is eligible to form stars
according to the Kennicutt-Schmidt Law when it satisfies a temperature
and density threshold. Our fiducial runs adopt $T < 15000$ K and
$n_{\rm th} > 10.3$ cm$^{-3}$. The stars feed energy back into the
interstellar medium (ISM) gas through blast-wave supernova feedback
(Stinson \etal 2006) and ionizing feedback from massive stars prior to
their explosion as supernovae, referred to as ``early stellar
feedback'' (Stinson \etal 2013).

In \gas\,, as in Stinson \etal (2013), the pre-SN feedback consists of
$\epsilon_{\rm ESF}=$10\% of the total stellar flux, 
$2\times10^{50}$ erg of thermal energy
per M$_\odot$ of the entire stellar population, being ejected from
stars into surrounding gas.  Because of the increased mixing in
\gastwo, the simulations required more stellar feedback to have their
star formation limited to the abundance matching value at the Milky
Way scale.  Thus, we set $\epsilon_{\rm ESF}$=13\%, which gives a
better match to Behroozi \etal (2013) abundance matching results.
Radiative cooling is left \emph{on} for the pre-SN feedback.
 
In the second, supernova, epoch of feedback, stars of mass $8$
M$_\odot < \Mstar < 40$ M$_\odot$ eject both energy and metals into
the interstellar medium gas surrounding the region where they formed.
Supernova feedback is implemented using the blastwave formalism
described in Stinson \etal (2006).  Since the gas receiving the energy
is dense, it would quickly be radiated away due to its efficient
cooling.  For this reason, cooling is delayed for particles inside the
blast region for $\sim30$ Myr.

\subsection{Derived galaxy and halo parameters}
In this section we describe how the various galaxy and halo parameters
we use in the rest of the paper are defined and measured.

Haloes in our zoom-in simulations were identified using the MPI+OpenMP
hybrid halo finder \texttt{AHF}\footnote{http://popia.ft.uam.es/AMIGA}
(Knollmann \& Knebe 2009; Gill \etal 2004). \texttt{AHF} locates local
over-densities in an adaptively smoothed density field as prospective
halo centers. The virial masses of the haloes are defined as the
masses within a sphere containing $\Delta=200$ times the cosmic
critical matter density.  The virial mass, and size are denoted
$M_{200}$ and $R_{200}$. The mass in stars, $M_{\rm star}$, is
measured within a sphere of radius, $r_{\rm gal}\equiv0.2
R_{200}$. The star formation rate, SFR, is measured as the mass of
stars formed inside $r_{\rm gal}$ over  the preceding 100 Myr.
These parameters of the main halo/galaxy are given in
Table~\ref{tab:parameters2}.  The simulation ID is named after the
redshift $z=0$ halo mass ($\Mvir/[\Msun/h]$) from the low resolution
dark matter-only simulation. The actual halo masses in the
hydrodynamic simulation may differ due to: baryonic mass loss;
differential evolution; and a different halo mass definition.

Edge-on and face-on images for a selection of 30 galaxies at $z=0$ are
shown in Figs.~\ref{fig:edgeon} \& \ref{fig:faceon}.  Each image is 50
kpc on a side and was created using the Monte Carlo radiative transfer
code {\sc{sunrise}} (Jonsson 2006). The image brightness and contrast
are scaled using arcsinh as described in Lupton \etal (2004).  The
image orientation was determined using the angular momentum vector of
the stars inside 5 kpc.   We see a variety of galaxy morphologies,
sizes, colors, and bulge-to-disc ratios.  The galaxies are ordered
according to their halo masses. The variation in morphology follows
the increase of halo mass.  The lowest mass galaxies are small blue
dots.  Higher mass halos gradually extend into peanut shapes before
filling out entire discs.  Once the galaxies become discs, their
stellar populations start to include more old, red stars, and exhibit
stronger absorption from dust lanes.  These discs then start to show
extended merger structures.  In some cases the disc has disappeared
and all that is left is a smaller, red spheroid that represents the
remnant of a merger.  Even though all the images are taken from the
$z=0$ outputs, the mass sequence seems to replicate the evolutionary
sequence of galaxies.

In addition to the main target halo each simulation contains several
smaller haloes within the high-res region, which we include in our
analysis. We verify with convergence tests (see \S\ref{sec:results} below) that these
poorer resolved haloes have similar global parameters to their higher
resolution counterparts.

\section{Results}
\label{sec:results}

\begin{figure*}
\centerline{
  \psfig{figure=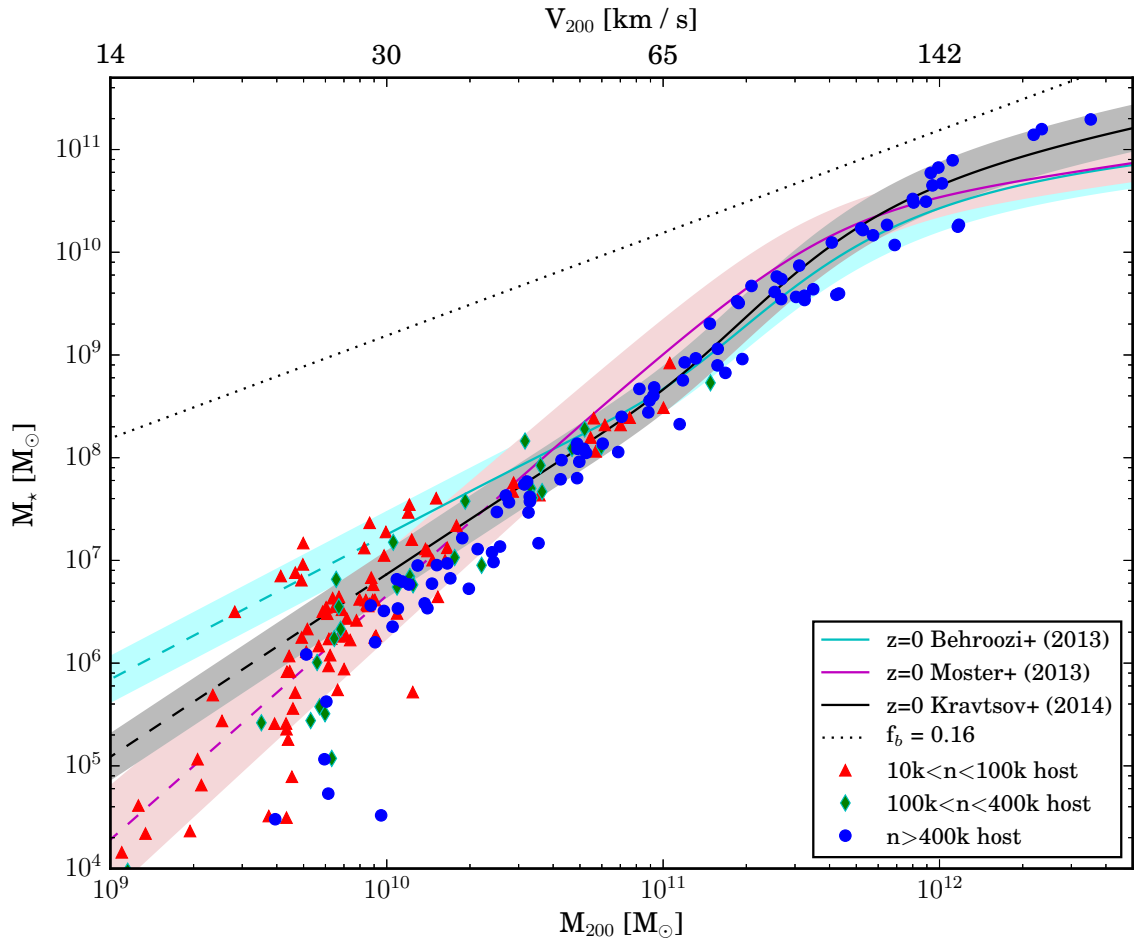,width=0.99\textwidth}
}
\caption{Stellar mass vs halo mass at redshift $z=0$ for main
  simulations (blue points) together with lower mass galaxies in the
  zoom-in region (green and red points). The solid black line and
  shaded region shows the relation from Kravtsov \etal (2014) derived
  using halo abundance matching. Our simulation matches this very
  well. The dashed lines show extrapolations of the abundance matching
  relations.  For reference, the dotted line shows the cosmic baryon
  fraction of mass associated with the dark matter halo, indicating
  that our simulations convert only a small fraction of the available
  gas into stars, as observed.}
\label{fig:ms_mh}
\end{figure*}

\begin{figure*}
\centerline{
  \psfig{figure=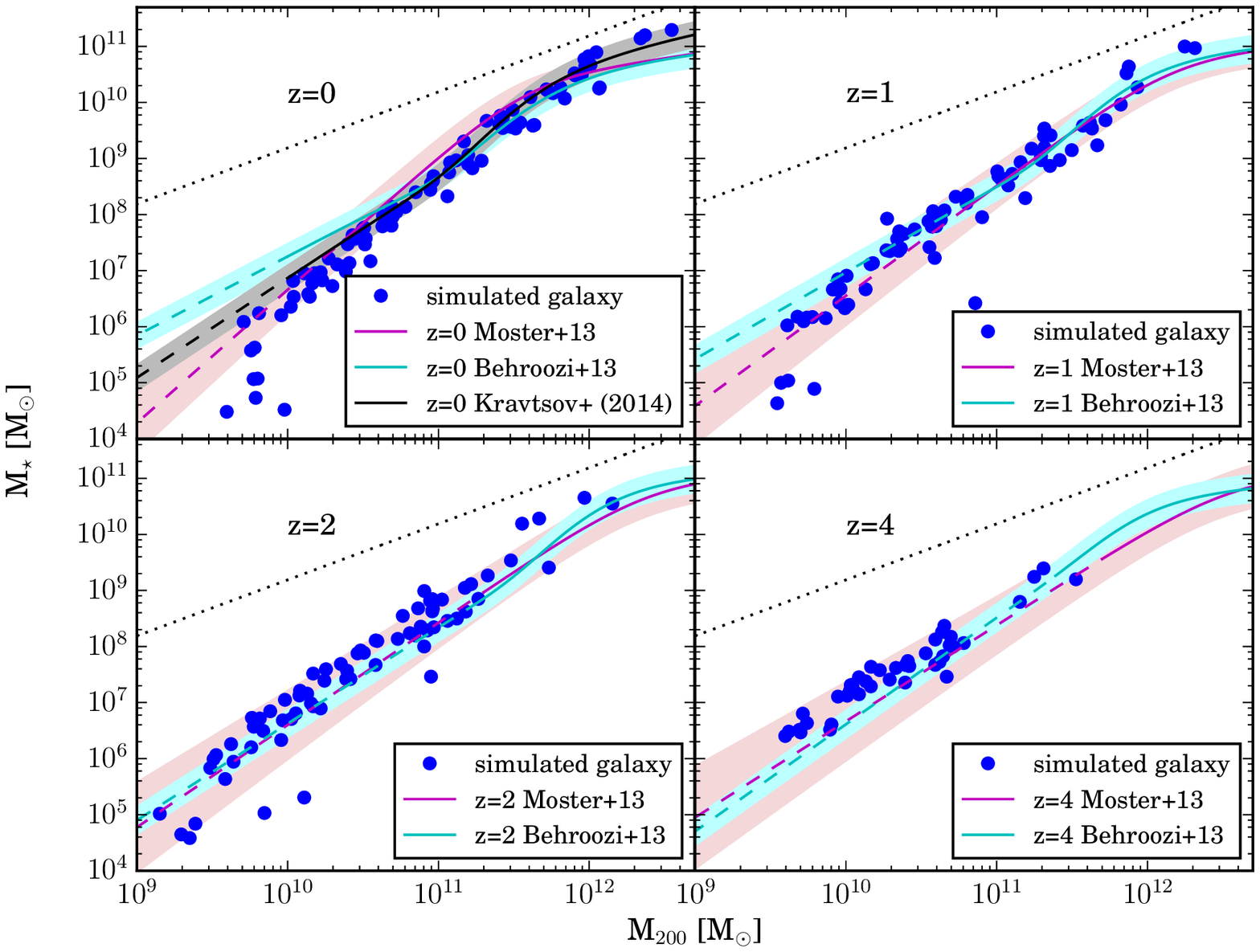,width=0.99\textwidth}
}
\caption{Evolution of the Stellar mass vs halo mass relation. Redshift
  $z=0$(top left), $z=1$(top right), $z=2$ (bottom left), and $z=4$
  (bottom right). Only the most massive halo in each zoom-in region
  are shown. There is good agreement between our simulations and
  constraints from halo abundance matching (Moster \etal 2013;
  Behroozi \etal 2013). The dashed lines show extrapolation of the
  relations into mass scales without observational constraints.
   For reference, the dotted line shows the cosmic baryon
  fraction of mass associated with the dark matter halo.}
\label{fig:ms_mh_z}
\end{figure*}

\subsection{Stellar mass vs halo mass}
\label{sec:mstar_mhalo}

Fig.~\ref{fig:ms_mh} shows the relation between stellar mass and
virial mass for our simulated central galaxies compared to results
from three different halo abundance matching measurements: Behroozi
\etal (2013), Moster \etal (2013), and Kravtsov \etal (2014).  We show
all three relations to represent the lingering uncertainty in the
abundance matching and halo occupation techniques.  Most of the
difference is apparent at low stellar masses, \Mstar $\lta 10^8
\Msun$.  That is the lower mass limit for the Sloan Digital Sky Survey
(SDSS, Blanton \etal 2005).  The smaller, deeper survey, GAMA (Driver
\etal 2012), has pushed the stellar mass function to lower masses, but
only the Kravtsov line uses data from the most up-to-date GAMA survey.
The lines at low masses are dashed indicating that they are
extrapolations of the higher mass results.   The Kravtsov relation
also differs from the other two at high masses because of how it takes
into account intra-cluster light, namely by using the Bernardi \etal
(2013) stellar mass function that extends surface brightness profiles
to larger radii. Our simulations agree well with each of these
relations, especially the most up-to-date relation from Kravtsov \etal
(2014) at high masses.  At low masses, the simulations most closely
follow the extrapolation of the Moster \etal (2013), which has the
steepest slope, $\Mstar \propto M_{200}^{2.4}$.

An additional systematic uncertainty to consider is due to form of the
stellar initial mass function (IMF).  The abundance matching results
use stellar masses assuming a universal Milky Way IMF (e.g., Chabrier
2003). A Milky Way IMF in spiral galaxies is consistent with dynamical
based mass-to-light ratios (Courteau \& Dutton 2015). However, there
is increasing evidence for ``heavier'' IMFs in massive elliptical
galaxies and the bulges of massive spiral galaxies (e.g., Conroy \&
van Dokkum 2012; Dutton \etal 2013a,b).  These studies suggest
  that for massive galaxies ($\Mstar \gta 10^{11}\Msun$) the 
  observed stellar masses may be underestimated by a factor of $\sim
2$ when assuming a Milky Way IMF and standard stellar population
  models.  Taking this into consideration nominally improves
the agreement of our most massive simulations with observations. 
  Note that formally we should use the same IMF in the observations
  and simulations. However, practically speaking the feedback
  efficiency parameters in the simulations can be simply rescaled.  In
  addition, there is enough observational uncertainty in the high mass
  end of the IMF, as well as the actual energy released per SN that
  this inconsistency is not significant.

The parameters of the feedback model are kept constant across the
entire mass range.  The parameters were determined based on a
comparison with abundance matching for one galaxy about the mass of
the Milky Way ($M_{200}\sim 10^{12} \Msun$) (see Stinson \etal
2013). The agreement between the simulated galaxies and the abundance
matching at a wide range of masses provides strong support  for
stellar feedback being the key ingredient that limits star formation
in galaxies lower mass than the Milky Way (see also Hopkins \etal
2014).

The points in Fig.~\ref{fig:ms_mh} are color coded  according to the
number of particles within the virial radius (dark, gas and
stars). The main haloes have between $4\times 10^5$ and $4\times 10^6$
particles. The halo mass range from $2\times 10^{10}$ to $10^{11}$
\Msun includes simulations with a wide range of resolutions (from 20
000 to more than 400 000  particles,  and dark matter particles' mass
is from  $6.4 \times 10^4$ to $1.7 \times 10^6 \Msun$).   There are no
systematic offsets indicating good convergence in stellar masses at an
order of magnitude fewer particles than our zoom-in target
galaxies. In addition, our central galaxies have a range of physical
spatial resolutions, with baryonic force softenings varying from  50 pc
to 400 pc.  This is encouraging as cosmological simulations do not
always converge when changing the physical resolution (see discussions
in Vogelsberger \etal 2014; Schaye \etal 2015).   

Below a halo mass of $M_{200} \sim 2\times 10^{10} \Msun$, there
starts to be a discrepancy between different resolutions.  The high
resolution simulations (blue points) contain less stars than lower
resolution galaxies.  There are two possibilities: Stars may form more
readily in low resolution galaxies or else their halo masses might be
reduced through enviromental processes such as tidal stripping.
While the lower resolution halos are outside of the main halos in the
$z=0$ output, it is possible that they have flown nearby the main
galaxy at some point in the past given their proximity to more massive
halos.  The low resolution halo could form a mass of stars according
to its earlier halo mass, make a close passage of the main halo, and
thus have its halo mass reduced while retaining its stellar mass (for
a study that shows that the outermost mass is stripped first, see
Chang \etal 2013).

Above a halo mass of $M_{200} \sim 2\times 10^{10} \Msun$ the scatter
in the relation is a constant and consistent with observational
constraints of $\sim 0.2$ dex (More \etal 2011; Reddick \etal
2013). Below a halo mass of $M_{200} \sim 2\times10^{10} \Msun$ the
scatter starts to increase.  More high resolution simulations are
needed to verify if this feature is due to the relatively low
resolution of these haloes or environmental effects.

A number of semi-analytic models have been able to produce galaxies
that follow the observed relation between stellar and halo mass at
redshift $z\sim 0$ (e.g., Bower \etal 2006; Somerville \etal 2008; Guo
\etal 2011). However, simultaneously reproducing the evolution has
been a challenge (Weinmann \etal 2012; Hopkins \etal 2014).
Fig.~\ref{fig:ms_mh_z} shows the evolution of the stellar mass vs halo
mass relation since redshift $z=4$ (a look back time of $\sim 12$
Gyr). Here we only show the most massive halo per simulation. Our
simulations show good agreement with the abundance matching relations
from Behroozi \etal (2013) and Moster \etal (2013).  Note that most of
the simulated galaxies fall in a mass range that can only be
extrapolated (dashed lines).  The range of masses directly constrained
by observations is shown as the solid lines. 

The agreement with the $\Mstar-M_{200}$ relation is not totally
unexpected,  as the feedback model employed here has been shown to
reproduce the time evolution of a single Milky Way mass galaxy through
the stellar mass vs halo mass plane (Stinson \etal 2013), as well as
the stellar to halo mass relation for galaxies of mass $10^{8.5} \lta
\Mstar \lta 10^{10.5}\Msun$ at redshifts $z>2$ using lower resolution
simulations in a cosmological volume of size $\sim 100$ Mpc (Kannan
\etal 2014).  Our study bridges the gap between these two works by
creating a large number of high-resolution simulations with realistic
stellar masses. 

Our simulations are not the first to replicate the $\Mstar-\Mhalo$
relation at $z=0$ or even at higher redshifts.  Aumer \etal (2013)
presented a sample of 16 galaxies with comparable resolution to
those presented here but on a smaller mass range ($10^{11}$ \Msun to 
$2\times10^{12}$ \Msun).  Those simulations showed a similar mass and 
redshift dependence as the simulations presented in our sample, even
though they tend to over produce stars at $z>4$ and under produce them
at $z<0.5$ in the low mass range, an effect we do not see in our
(even larger) sample.

Hopkins \etal (2014) presented a sample of only 7 galaxies with a
factor of $\sim 5$ higher mass resolution than ours (see
Fig.~\ref{fig:massres}), accompanied by several other galaxies that
were also simulated inside the zoom region.  Again, the results are
similar to what we find, although at all halo masses they overproduce
stellar masses at $z=2$ when compared with the Behroozi \etal (2013)
prediction and its extrapolation.  All three simulation suites were
run with different codes and stellar feedback recipes, but there are
common features to all of them.  1) They all include efficient
supernova feedback (at least $10^{51}$ erg returned to the ISM in all
cases).  2) The supernova feedback is deposited in the local
environment of where the stars formed, promptly, $\sim 3$ Myr  after
the stars formed.  3)  They all include some sort of feedback prior to
supernova (Aumer: strong, $\tau_{\rm IR}\sim 25$, radiation pressure;
Hopkins: radiation pressure, stellar winds, and photoionization; this
work: strong photoionization included as thermal energy).

\begin{figure*}
\centerline{
  \psfig{figure=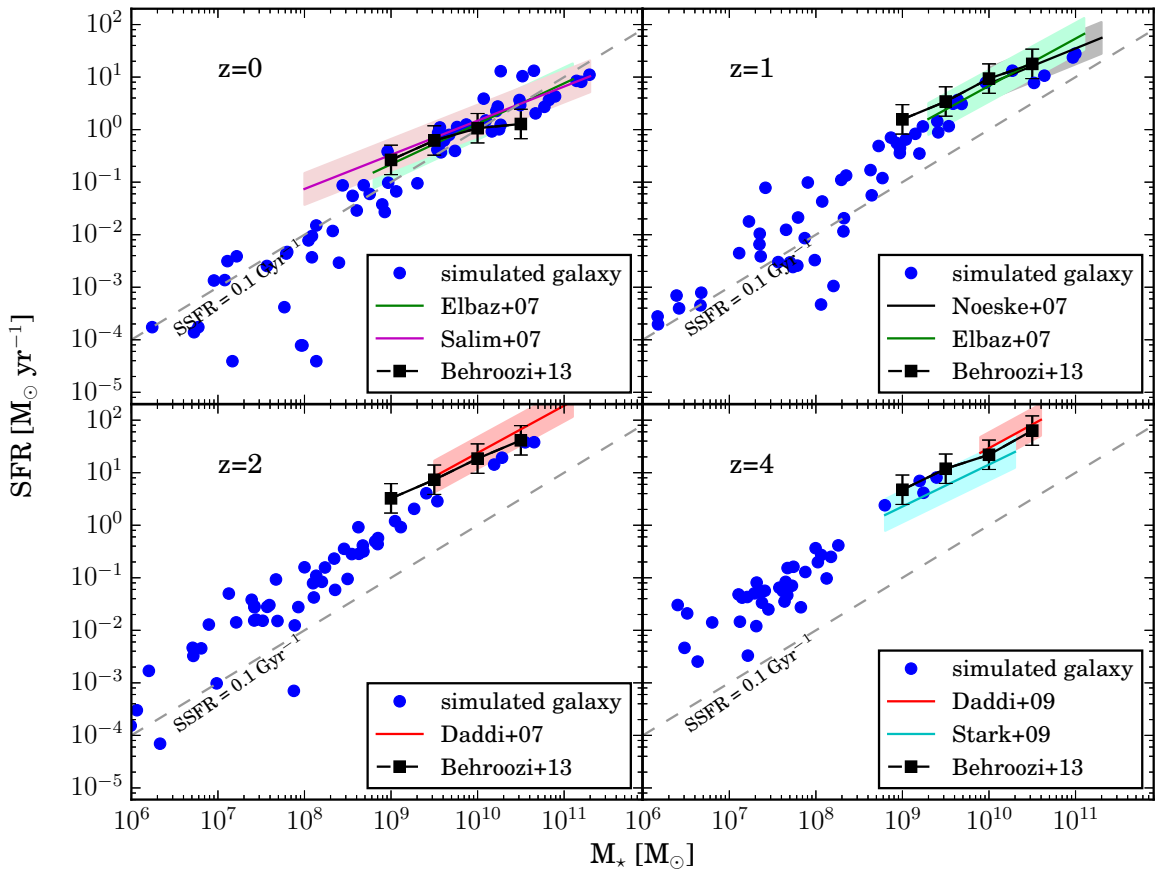,width=0.99\textwidth}
}
\caption{Evolution of the galaxy star formation rate vs stellar mass
  relation. NIHAO simulations are shown with blue points, observed
  relations by lines and shaded regions as indicated. For reference
  the dashed lines show a specific star formation rate,
  SSFR=SFR$/\Mstar=0.1$ Gyr$^{-1}$. The average SSFR of our
  simulations decline by a factor of $\sim 30$ from $z=4$ to the
  present day.}.
\label{fig:sfr_ms}
\end{figure*}

Schaye \etal (2015) shows that their simulated stellar mass function
follows observations at $z=0$, with an exception at the knee
that translates into a low peak efficiency of star formation 
around $\Mhalo\sim10^{12} \Msun$ (see their figure 8).
Those simulations invoke a simple prescription for stellar feedback
that injects slightly more thermal energy per supernova instantaneously
after the star particle is formed.  They add AGN feedback to 
their energy budget, which prevents over cooling in higher mass galaxies.  

\subsection{Star formation rate vs stellar mass}
\label{sec:sfr_mstar}

A consistency check on the stellar mass growth of our simulated
galaxies comes from the relation between star formation rate and
stellar mass, sometimes referred to as the ``star forming main sequence.''

Fig.~\ref{fig:sfr_ms} shows the evolution of the star forming main
sequence in our simulations compared with the observed relations (as
compiled by Dutton \etal 2010) at $z\sim 0$ (Elbaz \etal 2007, Salim
\etal 2007), $z\sim 1$ (Elbaz \etal 2007, Noeske \etal 2007), $z\sim
2$ (Daddi \etal 2007) and $z\sim4$ (Daddi \etal 2009, Stark \etal
2009), together with the compilation by Behroozi \etal (2013).
Broadly speaking, our simulations recover the order of magnitude
decline in SFR at fixed stellar mass observed since $z\sim 4$.In more detail, they
match the observations well at all but the lowest mass bins, where the
observations are more prone to selection biases.

The agreement at redshift $z=2$ is especially noteworthy, as a common
feature of galaxy formation models is to underpredict the observed
SFRs  by a factor of $\gta 2$  (e.g., Dutton \etal 2010; Yang \etal
2013; Romeo Velon\'a \etal 2013; Somerville \& Dav\'e 2014; Furlong
\etal 2015).  This feature of previous models is due to a strong
coupling between the cosmological gas accretion and star formation
rates (Weinmann \etal 2011) which leads to $SSFR\propto (1+z)^{2.25}$.
In our simulations this coupling is broken by our feedback model,
which effectively delays star formation at early times.  Other
solutions have included a modification to the stellar IMF (Dav\'e
2008) or an appeal to systematic biases in the observed  star
formation rates and stellar masses which could plausibly be at the
factor of $\sim 2$ level.

\begin{figure*}
\centerline{
\psfig{figure=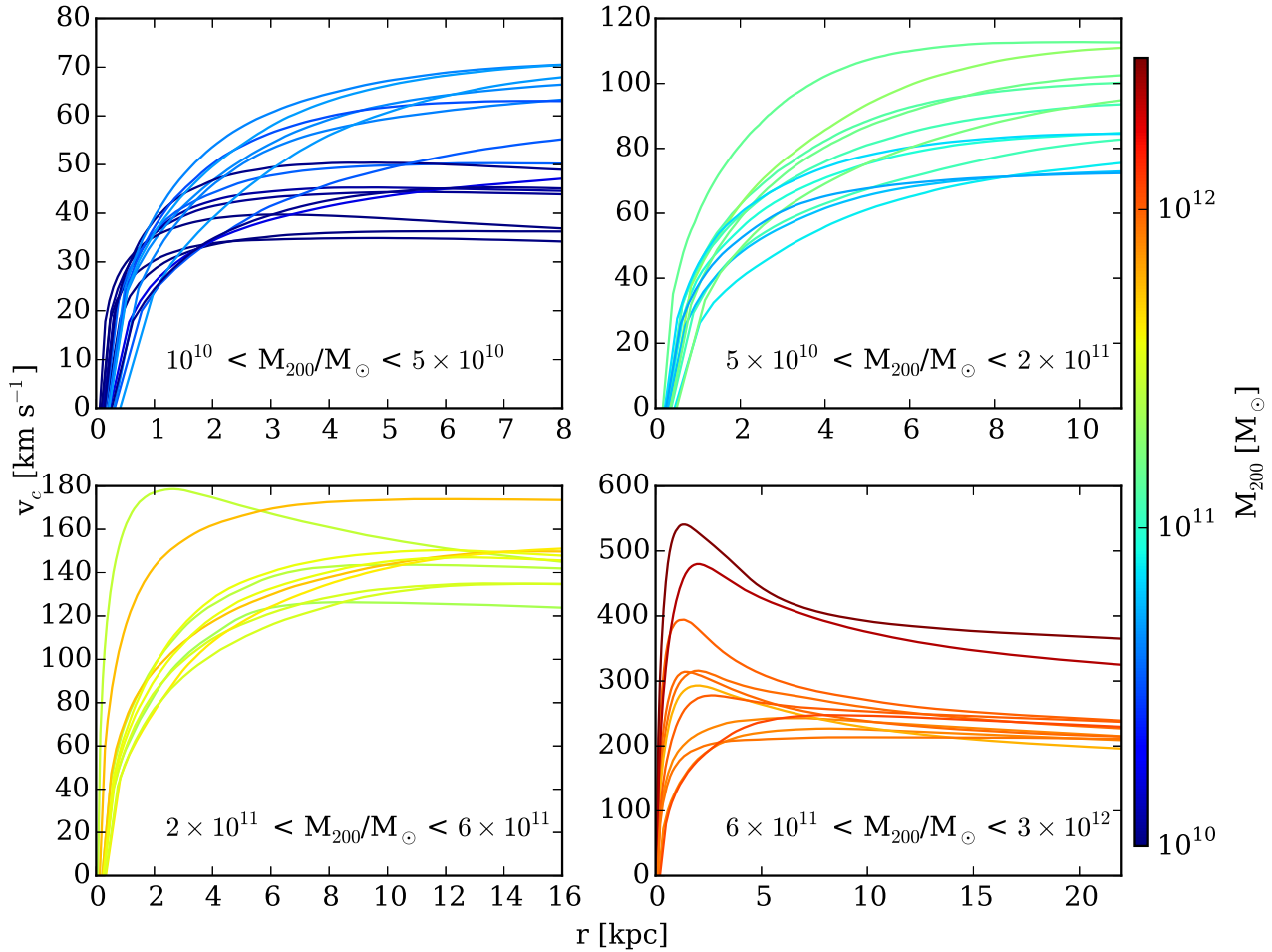,width=0.99\textwidth}
}
\caption{Rotation curves of the spherically averaged circular velocity
  $\sqrt{GM/r}$, split into four halo mass ranges, as indicated.}.
\label{fig:rcs}
\end{figure*}

\subsection{Rotation curves}

To this point, our analysis has focused exclusively on the mass of
stars formed in the simulated haloes.  Those comparisons show
that our simulations are successful according to that measure.
However, a fully successful galaxy formation model should 
also reproduce the mass \emph{distribution} of material in the
galaxies.  We will look into the matter distributions in more depth
in follow-up papers, but give a preliminary look at the rotation 
curves for the simulated galaxies here.

Fig. \ref{fig:rcs} shows the spherically averaged circular velocities
for all the galaxies in the NIHAO sample.  The galaxies have been
separated into four separate mass groups to limit the confusion.  The
circular velocity curves are colored by their halo mass.  The curves
show a gradual transition based on mass that represents a change in
the halo response that will be explored more explicitly in Dutton
\etal ({\em in prep}).  In general terms, the lowest mass halos,
$M_{200} < 2\times10^{10} \Msun$, have rotation curves that rise
relatively steeply.  As the masses increase, the rotation curves rise
more slowly.  Then, the galaxies cross a higher mass threshold around
$6\times10^{11} \Msun$ at which the galaxies revert to a steeply
rising rotation curve, which in some cases grows to have a high
central peak. Such centrally peaked rotation curves do exist in
observed massive spiral galaxies (e.g., Noordermeer \etal 2007; Dutton
\etal 2013a). However, we note that our simulations do not include any
form of AGN feedback. The central peaks in the rotation curves may
thus represent over cooling that would otherwise be heated by AGN or
another higher energy phenomenon.

\section{Summary}
\label{sec:sum}

We introduce a new, unmatched sample of high resolution hydrodynamic
simulations from the NIHAO project.  The galaxies are individually
simulated in their own zoom regions.  They span a wide range of halo
masses, $5\times 10^{9} < $M$_{200}/$M$_\odot < 3\times 10^{12}$ and
stellar masses $5\times 10^{4} < \Mstar/$M$_\odot < 2\times 10^{11}$.
Across this range of masses, the galaxies show a range of
morphologies, colors and sizes that correspond well with observed
galaxies.  The only energetic source term for the simulations comes
from stars: stellar feedback due to photoionization from hot,
young stars;  and the thermal energy from supernovae.  The stellar
masses of the simulations closely track the results from abundance
matching for stellar masses as  a function of halo mass across more
than three quarters of the life of the Universe.  That means that a
complete cosmological sample of  these galaxies would populate the
stellar mass function as observed.  The correspondence of the
simulations to observations provides a final proof that stellar
feedback is the chief piece of physics required to limit the
efficiency of star formation in galaxies less massive than the Milky
Way.  

\section*{Acknowledgments} 
The analysis was performed using the {\sc{pynbody}} package (Pontzen
\etal 2013), which had key contributions from Andrew Pontzen and  Rok
Ro\v{s}kar in addition to the authors.
The simulations were performed on the theo cluster of the
Max-Planck-Institut f\"ur Astronomie and the hydra cluster at the
Rechenzentrum in Garching; and the Milky Way supercomputer, funded by
the Deutsche Forschungsgemeinschaft (DFG) through Collaborative
Research Center (SFB 881) ``The Milky Way System'' (subproject Z2),
hosted and co-funded by the J\"ulich Supercomputing Center (JSC). We
greatly appreciate the contributions of all these computing
allocations.
AAD, AVM, and GSS acknowledge support through the
Sonderforschungsbereich SFB 881 ``The Milky Way System'' (subprojects
A1 and A2) of the German Research Foundation (DFG).
The authors acknowledge support from the MPG-CAS through the
partnership programme between the MPIA group lead by AVM and the PMO
group lead by XK. LW acknowledges support of the MPG-CAS student
programme. XK acknowledge the support from NSFC project
No.11333008 and the "Strategic Priority Research Program the Emergence
of Cosmological Structures" of the CAS(No.XD09010000).
GSS acknowledges funding  from the European Research Council under the
European Union's  Seventh Framework Programme (FP 7) ERC Grant
Agreement n. [321035].


\appendix
\section{PARAMETERS}

In this appendix we summarize all the galaxy and halo parameters
for main target haloes at redshift $z=0$ through this paper.
Table~\ref{tab:parameters2} contains the simulation ID, the number 
of total particles, dark matter particles and star particles, 
virial mass, stellar mass and star formation rate.


\begin{table*}
\caption{Galaxy and halo parameters for main target haloes at redshift $z=0$ }  
\label{tab:parameters2}
\begin{center}
\begin{tabular}{lrrrlll}
\hline
simulation ID & \# particles & \# DM particles & \# star particles & virial mass & stellar mass & star formation rate\\
              & $n_{\rm 200}$ & $n_{\rm DM}$ & $n_{\rm star}$ & ($M_{200}$/M$_\odot$) &($\Mstar$/M$_\odot$)  & (M$_\odot$ yr$^{-1}$)\\
\hline 
g3.54e09 & 1,200,374 & 1,157,470 & 85 & $3.95\times10^{9}$ & $3.02\times10^{4}$ & 0.00 \\ 
g4.36e09 & 1,278,589 & 1,167,760 & 79 & $9.54\times10^{9}$ & $3.30\times10^{4}$ & 0.00 \\ 
g4.99e09 & 320,446 & 295,765 & 538 & $5.72\times10^{9}$ & $3.76\times10^{5}$ & $5.8\times10^{-5}$ \\ 
g5.22e09 & 353,754 & 327,288 & 173 & $6.32\times10^{9}$ & $1.18\times10^{5}$ & $1.2\times10^{-5}$ \\ 
g5.41e09 & 466,210 & 443,516 & 2,956 & $5.11\times10^{9}$ & $1.21\times10^{6}$ & 0.00 \\ 
g5.59e09 & 358,332 & 335,527 & 2,520 & $6.46\times10^{9}$ & $1.74\times10^{6}$ & $1.7\times10^{-4}$ \\ 
g5.84e09 & 555,955 & 512,178 & 87 & $5.95\times10^{9}$ & $1.16\times10^{5}$ & $6.3\times10^{-5}$ \\ 
g7.05e09 & 584,878 & 544,306 & 3,249 & $1.05\times10^{10}$ & $2.27\times10^{6}$ & 0.00 \\ 
g7.34e09 & 559,724 & 521,731 & 331 & $6.05\times10^{9}$ & $4.22\times10^{5}$ & 0.00 \\ 
g9.26e09 & 581,361 & 527,275 & 132 & $6.14\times10^{9}$ & $5.37\times10^{4}$ & 0.00 \\ 
g1.09e10 & 690,285 & 546,318 & 9,204 & $1.09\times10^{10}$ & $6.55\times10^{6}$ & 0.00 \\ 
g1.18e10 & 628,317 & 565,451 & 4,944 & $1.10\times10^{10}$ & $3.41\times10^{6}$ & 0.00 \\ 
g1.23e10 & 515,485 & 468,423 & 2,314 & $9.08\times10^{9}$ & $1.60\times10^{6}$ & 0.00 \\ 
g1.44e10 & 1,159,093 & 834,135 & 9,656 & $1.70\times10^{10}$ & $6.69\times10^{6}$ & 0.00 \\ 
g1.47e10 & 862,592 & 784,867 & 13,043 & $1.52\times10^{10}$ & $9.00\times10^{6}$ & $1.3\times10^{-3}$ \\ 
g1.50e10 & 791,452 & 725,701 & 5,055 & $1.40\times10^{10}$ & $3.42\times10^{6}$ & 0.00 \\ 
g1.57e10 & 728,314 & 670,688 & 13,027 & $1.29\times10^{10}$ & $8.93\times10^{6}$ & 0.00 \\ 
g1.88e10 & 1,063,823 & 971,852 & 23,864 & $1.88\times10^{10}$ & $1.65\times10^{7}$ & $3.9\times10^{-3}$ \\ 
g1.89e10 & 1,401,592 & 1,060,626 & 17,813 & $2.13\times10^{10}$ & $1.29\times10^{7}$ & $3.1\times10^{-3}$ \\ 
g1.90e10 & 1,457,806 & 1,222,038 & 17,279 & $2.40\times10^{10}$ & $1.20\times10^{7}$ & $1.4\times10^{-3}$ \\ 
g1.92e10 & 1,272,888 & 990,462 & 8,255 & $1.98\times10^{10}$ & $5.30\times10^{6}$ & $1.4\times10^{-4}$ \\ 
g1.95e10 & 794,621 & 704,686 & 5,488 & $1.37\times10^{10}$ & $3.82\times10^{6}$ & 0.00 \\ 
g2.09e10 & 1,018,898 & 839,068 & 13,600 & $1.66\times10^{10}$ & $9.35\times10^{6}$ & 0.00 \\ 
g2.34e10 & 1,500,299 & 1,319,369 & 19,729 & $2.57\times10^{10}$ & $1.36\times10^{7}$ & 0.00 \\ 
g2.37e10 & 1,503,953 & 1,232,912 & 25,627 & $2.43\times10^{10}$ & $9.65\times10^{6}$ & 0.00 \\ 
g2.39e10 & 869,294 & 743,922 & 8,601 & $1.46\times10^{10}$ & $5.94\times10^{6}$ & $1.7\times10^{-4}$ \\ 
g2.63e10 & 458,723 & 414,291 & 18,388 & $2.70\times10^{10}$ & $4.28\times10^{7}$ & 0.00 \\ 
g2.64e10 & 2,340,543 & 1,579,110 & 43,275 & $3.26\times10^{10}$ & $2.93\times10^{7}$ & 0.00 \\ 
g2.80e10 & 521,957 & 413,400 & 15,628 & $2.77\times10^{10}$ & $3.68\times10^{7}$ & $2.5\times10^{-3}$ \\ 
g2.83e10 & 483,150 & 370,911 & 13,428 & $2.50\times10^{10}$ & $2.96\times10^{7}$ & 0.00 \\ 
g2.94e10 & 1,908,993 & 1,657,184 & 85,637 & $3.22\times10^{10}$ & $5.86\times10^{7}$ & $4.2\times10^{-4}$ \\ 
g3.19e10 & 677,287 & 525,030 & 10,679 & $3.54\times10^{10}$ & $1.47\times10^{7}$ & $3.9\times10^{-5}$ \\ 
g3.44e10 & 1,077,981 & 694,131 & 26,561 & $4.88\times10^{10}$ & $6.32\times10^{7}$ & $4.7\times10^{-3}$ \\ 
g3.67e10 & 547,857 & 480,668 & 23,584 & $3.15\times10^{10}$ & $5.49\times10^{7}$ & 0.00 \\ 
g3.93e10 & 682,659 & 477,948 & 16,007 & $3.30\times10^{10}$ & $3.75\times10^{7}$ & 0.00 \\ 
g4.27e10 & 885,293 & 616,773 & 26,191 & $4.25\times10^{10}$ & $6.16\times10^{7}$ & $4.4\times10^{-3}$ \\ 
g4.48e10 & 1,191,821 & 895,736 & 58,334 & $6.04\times10^{10}$ & $1.37\times10^{8}$ & $3.9\times10^{-5}$ \\ 
g4.86e10 & 1,051,559 & 757,795 & 52,672 & $5.16\times10^{10}$ & $1.22\times10^{8}$ & $3.7\times10^{-3}$ \\ 
g4.94e10 & 984,362 & 791,322 & 48,035 & $5.26\times10^{10}$ & $1.11\times10^{8}$ & $7.8\times10^{-3}$ \\ 
g4.99e10 & 940,085 & 732,808 & 52,511 & $4.90\times10^{10}$ & $1.22\times10^{8}$ & $9.4\times10^{-3}$ \\ 
g5.05e10 & 777,412 & 650,173 & 40,840 & $4.29\times10^{10}$ & $9.47\times10^{7}$ & $7.8\times10^{-5}$ \\ 
g6.12e10 & 986,798 & 733,881 & 39,024 & $4.97\times10^{10}$ & $9.13\times10^{7}$ & $7.8\times10^{-5}$ \\ 
g6.37e10 & 2,621,929 & 1,617,864 & 95,020 & $1.15\times10^{11}$ & $2.12\times10^{8}$ & $1.2\times10^{-2}$ \\ 
g6.77e10 & 2,118,102 & 1,333,784 & 204,017 & $9.28\times10^{10}$ & $4.84\times10^{8}$ & $8.7\times10^{-2}$ \\ 
g6.91e10 & 1,331,679 & 1,069,321 & 107,120 & $7.08\times10^{10}$ & $2.50\times10^{8}$ & $2.9\times10^{-3}$ \\ 
g7.12e10 & 1,009,755 & 715,315 & 58,766 & $4.88\times10^{10}$ & $1.37\times10^{8}$ & $1.5\times10^{-2}$ \\ 
g8.89e10 & 591,924 & 397,347 & 50,174 & $9.22\times10^{10}$ & $4.02\times10^{8}$ & $2.9\times10^{-2}$ \\ 
g9.59e10 & 614,423 & 368,306 & 34,468 & $8.84\times10^{10}$ & $2.76\times10^{8}$ & $8.7\times10^{-2}$ \\ 
g1.05e11 & 775,657 & 505,507 & 70,478 & $1.18\times10^{11}$ & $5.67\times10^{8}$ & $6.0\times10^{-2}$ \\ 
g1.08e11 & 785,537 & 520,108 & 108,025 & $1.20\times10^{11}$ & $8.47\times10^{8}$ & $2.7\times10^{-2}$ \\ 
g1.37e11 & 1,014,522 & 653,073 & 252,530 & $1.48\times10^{11}$ & $2.02\times10^{9}$ & $9.5\times10^{-2}$ \\ 
g1.52e11 & 1,128,519 & 657,385 & 112,285 & $1.57\times10^{11}$ & $7.92\times10^{8}$ & $3.8\times10^{-2}$ \\ 
g1.57e11 & 1,076,183 & 677,124 & 143,921 & $1.58\times10^{11}$ & $1.15\times10^{9}$ & $6.7\times10^{-2}$ \\ 
g1.59e11 & 1,223,754 & 691,453 & 86,944 & $1.68\times10^{11}$ & $6.69\times10^{8}$ & $7.9\times10^{-4}$ \\ 
g1.64e11 & 1,468,563 & 786,245 & 111,008 & $1.93\times10^{11}$ & $9.13\times10^{8}$ & 0.38 \\ 
g2.04e11 & 1,722,119 & 901,588 & 573,355 & $2.09\times10^{11}$ & $4.70\times10^{9}$ & 0.79 \\ 
g2.19e11 & 920,447 & 557,247 & 113,958 & $1.31\times10^{11}$ & $9.27\times10^{8}$ & $9.8\times10^{-2}$ \\ 
g2.39e11 & 2,133,166 & 1,109,605 & 698,696 & $2.58\times10^{11}$ & $5.80\times10^{9}$ & 1.12 \\ 
g2.41e11 & 1,939,348 & 1,092,825 & 501,991 & $2.53\times10^{11}$ & $4.10\times10^{9}$ & 0.62 \\ 
g2.42e11 & 2,073,620 & 1,175,947 & 689,091 & $2.68\times10^{11}$ & $5.48\times10^{9}$ & 0.39 \\ 
g2.54e11 & 2,004,595 & 1,144,673 & 424,390 & $2.67\times10^{11}$ & $3.50\times10^{9}$ & 0.87 \\ 

 \hline
\end{tabular}
\end{center}
\end{table*}

\setcounter{table}{1}
\begin{table*}
\caption{cont'd Galaxy and halo parameters for main target haloes at redshift $z=0$ }  
\label{tab:parameters3}
\begin{center}
\begin{tabular}{lrrrlll}
  \hline
  simulation ID & \# particles & \# DM particles & \# star particles & virial mass & stellar mass & star formation rate\\
              & $n_{\rm 200}$ & $n_{\rm DM}$ & $n_{\rm star}$ & ($M_{200}$/M$_\odot$) &($\Mstar$/M$_\odot$)  & (M$_\odot$ yr$^{-1}$)\\
\hline 
g3.06e11 & 2,688,990 & 1,323,580 & 906,552 & $3.11\times10^{11}$ & $7.42\times10^{9}$ & 1.25 \\ 
g3.21e11 & 2,342,924 & 1,272,210 & 448,644 & $3.03\times10^{11}$ & $3.67\times10^{9}$ & 1.10 \\ 
g3.23e11 & 631,510 & 371,614 & 44,745 & $8.93\times10^{10}$ & $3.60\times10^{8}$ & $5.5\times10^{-2}$ \\ 
g3.49e11 & 3,276,200 & 1,815,975 & 539,486 & $4.33\times10^{11}$ & $3.97\times10^{9}$ & 0.61 \\ 
g3.55e11 & 3,241,957 & 1,760,597 & 480,380 & $4.23\times10^{11}$ & $3.85\times10^{9}$ & 0.66 \\ 
g3.59e11 & 2,762,536 & 1,459,228 & 534,349 & $3.49\times10^{11}$ & $4.36\times10^{9}$ & 0.74 \\ 
g3.71e11 & 4,015,748 & 1,691,651 & 1,535,547 & $4.08\times10^{11}$ & $1.24\times10^{10}$ & 1.49 \\ 
g4.90e11 & 2,452,320 & 1,369,494 & 428,202 & $3.25\times10^{11}$ & $3.43\times10^{9}$ & 0.42 \\ 
g5.02e11 & 5,258,849 & 2,408,247 & 1,821,403 & $5.75\times10^{11}$ & $1.46\times10^{10}$ & 0.92 \\ 
g5.31e11 & 3,534,992 & 2,204,411 & 559,381 & $5.28\times10^{11}$ & $1.64\times10^{10}$ & 2.26 \\ 
g5.36e11 & 4,505,028 & 2,889,033 & 445,661 & $6.88\times10^{11}$ & $1.18\times10^{10}$ & 3.88 \\ 
g5.38e11 & 4,187,358 & 2,726,961 & 648,971 & $6.46\times10^{11}$ & $1.85\times10^{10}$ & 1.24 \\ 
g5.46e11 & 2,459,236 & 1,374,445 & 470,795 & $3.25\times10^{11}$ & $3.78\times10^{9}$ & 0.37 \\ 
g5.55e11 & 5,020,346 & 2,200,598 & 2,075,884 & $5.20\times10^{11}$ & $1.71\times10^{10}$ & 2.75 \\ 
g6.96e11 & 1,094,722 & 412,962 & 498,881 & $8.00\times10^{11}$ & $3.30\times10^{10}$ & 10.40 \\ 
g7.08e11 & 1,098,831 & 412,287 & 479,668 & $8.06\times10^{11}$ & $3.05\times10^{10}$ & 3.66 \\ 
g7.44e11 & 1,294,128 & 596,875 & 283,637 & $1.18\times10^{12}$ & $1.85\times10^{10}$ & 12.95 \\ 
g7.55e11 & 1,185,149 & 455,930 & 483,752 & $8.93\times10^{11}$ & $3.11\times10^{10}$ & 2.92 \\ 
g7.66e11 & 1,528,867 & 478,007 & 911,988 & $9.30\times10^{11}$ & $5.92\times10^{10}$ & 2.71 \\ 
g8.06e11 & 1,366,038 & 481,349 & 665,796 & $9.43\times10^{11}$ & $4.48\times10^{10}$ & 13.19 \\ 
g8.13e11 & 1,704,130 & 505,171 & 1,043,618 & $9.91\times10^{11}$ & $6.68\times10^{10}$ & 3.67 \\ 
g8.26e11 & 1,513,265 & 518,236 & 739,749 & $1.02\times10^{12}$ & $4.68\times10^{10}$ & 2.04 \\ 
g8.28e11 & 1,276,937 & 591,330 & 275,702 & $1.16\times10^{12}$ & $1.77\times10^{10}$ & 1.01 \\ 
g1.12e12 & 1,977,112 & 564,785 & 1,222,359 & $1.12\times10^{12}$ & $7.85\times10^{10}$ & 4.29 \\ 
g1.77e12 & 3,668,318 & 1,117,439 & 2,173,715 & $2.19\times10^{12}$ & $1.39\times10^{11}$ & 8.47 \\ 
g1.92e12 & 4,018,048 & 1,200,667 & 2,467,821 & $2.34\times10^{12}$ & $1.58\times10^{11}$ & 8.13 \\ 
g2.79e12 & 5,598,386 & 1,800,095 & 3,099,962 & $3.53\times10^{12}$ & $1.96\times10^{11}$ & 11.09 \\ 

 \hline
\end{tabular}
\end{center}
\end{table*}

\label{lastpage}
\end{document}